\documentclass[conference]{IEEEtran}
\IEEEoverridecommandlockouts

\usepackage{amsmath,amssymb,amsfonts}
\usepackage{graphicx}
\usepackage{adjustbox}
\usepackage{import}
\usepackage{cite}
\usepackage{wrapfig}
\usepackage{algorithm}
\usepackage{algpseudocode}
\usepackage{textcomp}
\usepackage{xcolor}
\usepackage{svg}
\usepackage{subfig}
\usepackage{soul}
\usepackage{multirow}
\usepackage{booktabs}
\usepackage{enumitem}
\usepackage{hyperref}
\usepackage{stfloats} 
\setlist[itemize]{noitemsep, topsep=0pt}

\long\def\/*#1*/{}

\newcommand\asp{\texttt{asp}}

\def\BibTeX{{\rm B\kern-.05em{\sc i\kern-.025em b}\kern-.08em
    T\kern-.1667em\lower.7ex\hbox{E}\kern-.125emX}}

\title{High-Resolution Range Profile Classifiers Require Aspect-Angle Awareness%
\\
\thanks{Submitted at Eusipco26. This work was performed using HPC resources from GENCI-IDRIS (Grant 2025-AD011014422R2).}}
\author{\IEEEauthorblockN{Edwyn Brient}
\IEEEauthorblockA{\textit{STIM, Mines Paris, PSL University}\\
\textit{ARC, Thales Land and Air Systems}\\
Fontainebleau/Limours, France\\ edwyn.brient@minesparis.psl.eu}
\and
\IEEEauthorblockN{Santiago Velasco-Forero}
\IEEEauthorblockA{\textit{STIM, Mines Paris, PSL University}\\
Fontainebleau, France\\ santiago.velasco@minesparis.psl.eu}
\and
\IEEEauthorblockN{Rami Kassab}
\IEEEauthorblockA{\textit{ARC, Thales Land and Air Systems}\\
Limours, France\\ rami.kassab@thalesgroup.com}}

\begin{document}
%
\maketitle
\begin{abstract} 
We revisit High-Resolution Range Profile (HRRP) classification with aspect-angle conditioning. While
prior work often assumes that aspect-angle information is incomplete during training or unavailable at
inference, we study a setting where angles are available for all training samples and explicitly
provided to the classifier. Using three datasets and a broad range of conditioning strategies and model
architectures, we show that both single-profile and sequential classifiers benefit consistently from
aspect-angle awareness, with an average accuracy gain of about 7\% and improvements of up to 10\%,
depending on the model and dataset. In practice, aspect angles are not directly measured and must be
estimated. We show that a causal Kalman filter can estimate them online with a median error of 5°, and
that training and inference with estimated angles preserves most of the gains, supporting the proposed
approach in realistic conditions.
\end{abstract}


\section{Introduction}
\label{sec:intro}
Recent advances in radar resolution have enabled target representations to evolve from isolated point 
detections to two-dimensional response maps. However, because of the high scan rate and the large surveillance area,
processing such high-resolution grids remains computationally demanding. Consequently, long-range 
radars commonly compress the target response into a one-dimensional High-Resolution Range Profile 
(HRRP) by projecting the received echoes onto the radar line of sight (LOS). This dimensionality reduction 
retains the dominant structural characteristics of the target while significantly decreasing data 
volume, thereby facilitating real-time processing at the expense of fine-scale spatial details.

The interest in HRRP data for radar automatic target recognition (RATR) has grown significantly over
the past decade, driven by the need for rapid onboard classification in dynamic environments. Numerous
studies have demonstrated the effectiveness of machine learning methods applied to HRRP data for target
classification, both using a single profile as input \cite{Wan2019ConvolutionalNN, bauw2020unsupervised, Sun,
hrrp_clf1} and using sequences of profiles \cite{hrrp_seq1, hrrp_seq2, hrrp_seq3}. However, the strong
sensitivity of HRRP signals to the aspect angle has only recently been explicitly emphasized. Existing works
addressing the impact of aspect angle follow diverse paradigms, including data generation and evaluation
metrics \cite{multi_aspect_gen, mfn}, domain adaptation \cite{domain_adapt_hrrp, domain_adapt_hrrp2}, and
few-shot classification \cite{constrastive_learning_ma}.

We study the benefits of aspect-angle awareness for HRRP classification under realistic angle estimates.
Previous works generally assume that aspect-angle information is incomplete in training datasets or
unavailable during inference. In contrast, we consider a training scenario in which aspect angles are
available for all training samples and analyze the impact of explicitly providing this information to
the classifier. Using an extensive set of conditioning mechanisms, model architectures, and three datasets, we 
show that both single-profile classifiers and sequential classifiers that incorporate aspect-angle information 
outperform their angle-unaware counterparts. However, in real-world scenarios, aspect angles are not 
directly measured during acquisition and must be estimated. We first demonstrate that a Kalman filter 
can estimate the aspect angle accurately in real time. Our final experiments involve training models with 
estimated aspect angles without degrading classification performance, demonstrating the feasibility of our approach.
Some of the data and code used in this work are available at \href{https://github.com/EdwynBrient/HRRPclf-req-angles}{https://github.com/EdwynBrient/HRRPclf-req-angles}.
Our contributions can be summarized as follows:

\begin{itemize}
      \item \textbf{Aspect angle awareness for HRRP classification.} Using both two measured HRRP ship datasets and a MSTAR-derived
      HRRP dataset, we demonstrate that single-data classifiers and sequential classifiers that are
      aware of the aspect angle outperform those that are not.

      \item \textbf{Aspect-angle estimation.} We show that a causal Kalman filter estimates ship aspect angles online from
      AIS kinematics with low error.

      \item \textbf{Practical angle-aware classification.} We demonstrate that conditioning on estimated angles achieves
      strong performance, supporting deployment when angles are not directly measured.
\end{itemize}

\section{High-Resolution Range Profile Background}
\subsection{HRRP Data}
\label{subsec:HRRPdata}

A radar measures the backscattered returns of its transmitted waveform and,
after standard front-end processing \cite{richards2005fundamentals}, organizes
them into a polar map of radar cross section (RCS) values $\sigma(r,\theta)$,
indexed by range $r$ (distance) and azimuth $\theta$ (bearing). A
one-dimensional range profile is obtained by aggregating RCS values over the
azimuth span of the detection cone $[\Theta, \Theta+\Delta\Theta]$:

\begin{equation}
\mathrm{HRRP}(r_i)=\sum_{\theta_j\in[\Theta,\Theta+\Delta\Theta]}\sigma(r_i,\theta_j).
\end{equation}
The range-bin spacing defines the range resolution $\Delta r$.

The HRRP shape depends on the acquisition geometry, primarily governed by the
\emph{aspect angle} $\asp$ and the \emph{depression angle}. The aspect angle is
defined as the relative orientation between the target heading $hdg$ and the
radar azimuth $\theta$, i.e., $\asp = hdg - \theta$. The depression angle is
the elevation of the radar line of sight (LOS) with respect to the horizontal
plane.

\begin{figure}[!htbp]
    \centering
    \includegraphics[width=0.78\linewidth]{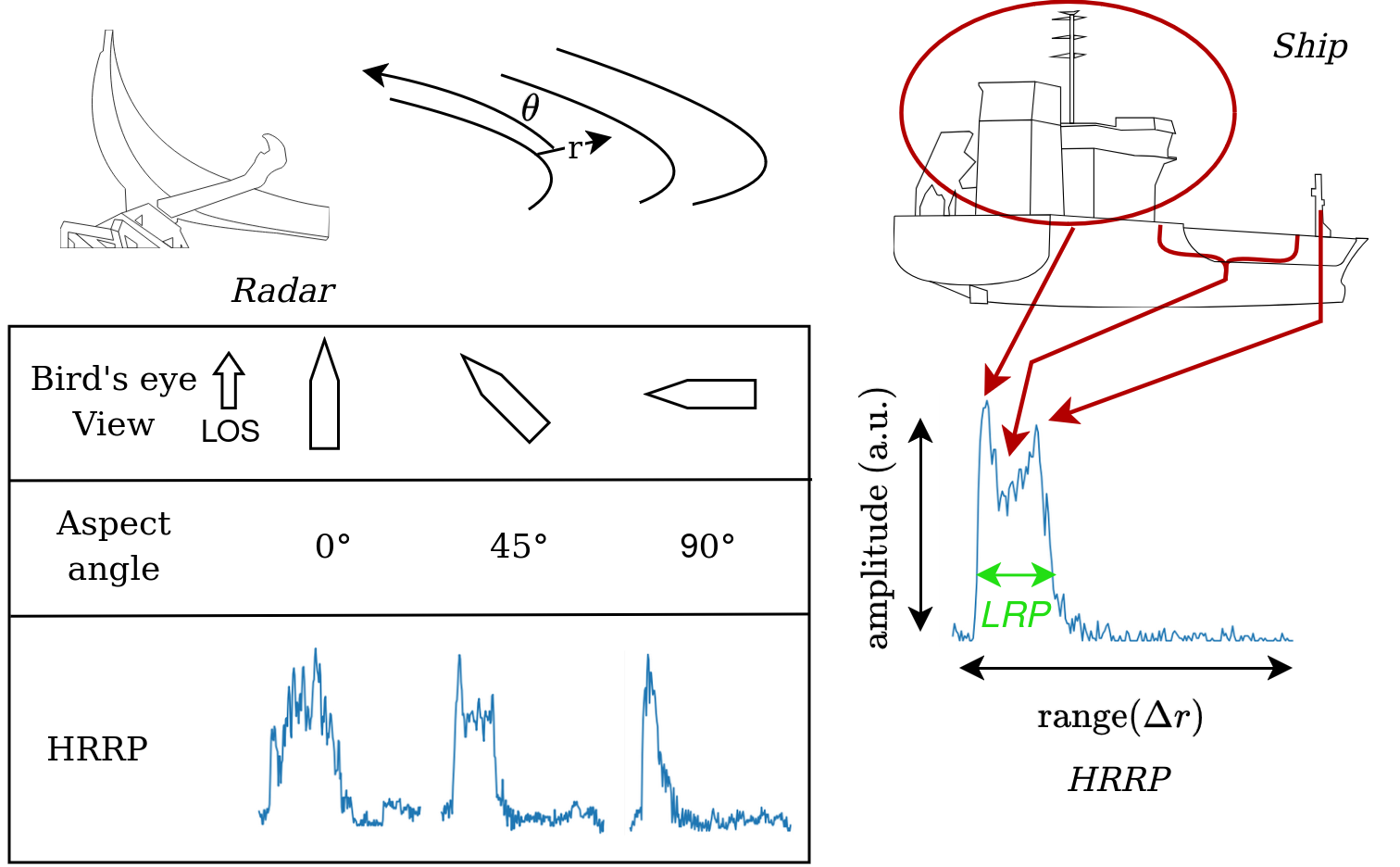}
    \caption{\textit{HRRP structure and aspect-angle dependence:} an HRRP sums the
    echoes of dominant scatterers within each range cell; changing $\asp$ alters
    the coarse-scale signature.}
    \label{fig:hrrp_def_sch}
    \vspace{-10pt}
\end{figure}

Because a 2D scattering distribution is projected onto a 1D profile, HRRPs are
not unique: distinct targets can yield similar profiles under certain viewing
conditions. In this work, we show that providing aspect-angle information helps
reduce this ambiguity and improves classification performance.

\subsection{Aspect Angle and HRRP Geometry}

\begin{wrapfigure}[12]{r}{0.5\linewidth}
    \centering
    \vspace{-6pt}
    \includegraphics[width=\linewidth]{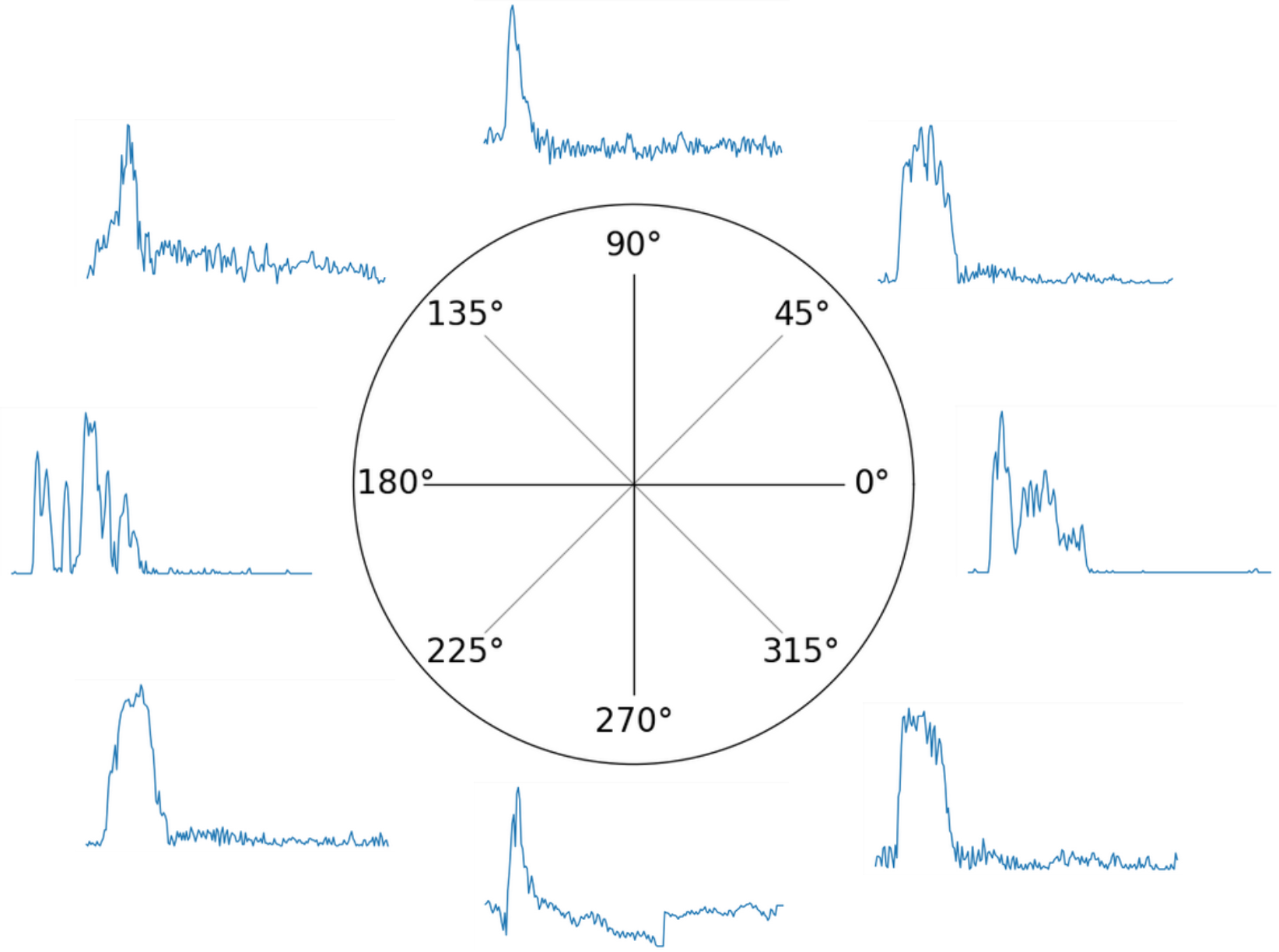}
    \caption{HRRP of a ship at multiple aspect angles.}
    \label{fig:trigo_hrrp}
    \vspace{-10pt}
\end{wrapfigure}

At fine scale, HRRP signatures are not expected to be $\pi$-invariant because
dominant scatterers and occlusion/shadowing depend on the viewing direction.
However, a coarser geometric cue is often close to $\pi$-periodic: up to a few
strong reflections, the occupied extent along range is mainly driven by the
target length projected onto the LOS. This is visible in
Fig.~\ref{fig:trigo_hrrp}, where the overall support remains similar for angles
separated by $\pi$ (with noticeable deviations around $135^\circ$ and
$315^\circ$).

\begin{figure}[!htbp]
    \centering
    \includegraphics[width=0.7\linewidth]{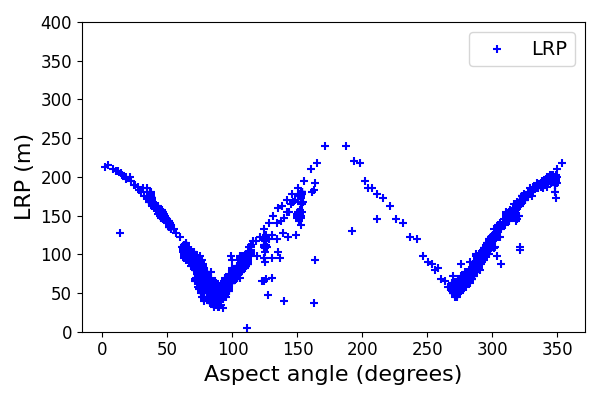}
    \caption{Length on Range Profile (LRP) \cite{mfn} across aspect angles.}
    \label{fig:lrp_losp}
    \vspace{-10pt}
\end{figure}

The \emph{Length on Range Profile} (LRP) \cite{mfn} summarizes this support by
measuring the range extent of the target response, i.e., an estimate of the
LOS-projected length. Fig.~\ref{fig:lrp_losp} illustrates its approximate
$\pi$-periodicity.

\section{Methods}

\subsection{Models}
We evaluate aspect-angle conditioning in two classification settings. In the
\emph{single-view} setting, the input is a single HRRP, and the model predicts
its class label from that profile alone. In the \emph{multi-view} setting, the
input is a temporally ordered sequence of HRRPs acquired along the same target
trajectory; the model aggregates per-profile features to produce a single
prediction for the sequence.

All models follow a two-stage design: a feature extractor followed by a
classifier head. The extractor maps an input HRRP to a compact latent vector,
which is then converted into class logits by the head. For single-view
experiments, we consider three extractor families: (i) a ResNet-style 1D
backbone \cite{resnet}, (ii) a standard convolutional network, and (iii) a
multilayer perceptron (MLP). In the multi-view setting, we use the ResNet
backbone as the per-profile extractor, as it consistently performed best in our
preliminary single-view experiments. To keep the multi-view study focused on
temporal aggregation and angle conditioning, we fix this backbone and only vary
the sequence model (LSTM, GRU, or Transformer).

All conditioning mechanisms (Sec.~\ref{subsec:conditioning_methods}) are
implemented within the feature extractor. Aspect-angle information is injected
at multiple depths: after each residual block (ResNet), after each convolutional
block (CNN), and after each hidden layer (MLP). At each injection point, we use
a dedicated predictor $f_\theta$ that matches the current channel dimension and
apply conditioning before the nonlinearity. To account for class imbalance
(Fig.~\ref{fig:mmsi_compact}), we train all models end-to-end with a weighted

cross-entropy loss (inverse class frequencies). In addition to overall accuracy, 
we report the macro-averaged F1 score, which assigns equal weight to each class. 
For a given class, precision (P), recall (R), and the F1 score are defined:
\begin{equation}
\mathrm{P}=\frac{\mathrm{TP}}{\mathrm{TP}+\mathrm{FP}},\quad
\mathrm{R}=\frac{\mathrm{TP}}{\mathrm{TP}+\mathrm{FN}},\quad
\mathrm{F1}=\frac{2\,\mathrm{P}\,\mathrm{R}}{\mathrm{P}+\mathrm{R}}.
\end{equation}
where TP, FP, and FN denote true positives, false positives, and false negatives, 
respectively. The macro-F1 score is computed as the arithmetic mean of the per-class 
F1 scores giving equal importance to each class regardless of its frequency.

We intentionally rely on standard backbones (ResNet-1D, ConvNet, and MLP) and focus on quantifying the
impact of aspect-angle conditioning and multi-view aggregation on HRRP recognition. Architectural and
training hyperparameters are kept fixed across conditioning methods for fair comparison and are
provided in the accompanying code repository.

\subsection{Aspect-Angle Estimation}
\label{subsec:aspect_angle_estimation}
In operational settings, aspect angles are not directly measured and must be
estimated from kinematic information. We apply a Kalman filter \cite{welch1995introduction} to denoise
position measurements and obtain a smoothed state estimate at each time step,
$(x_t, y_t, \dot{x}_t, \dot{y}_t)$, where $(x_t,y_t)$ denotes the target
position and $(\dot{x}_t,\dot{y}_t)$ its velocity. The target heading is then
computed from the predicted velocity:
{\setlength{\abovedisplayskip}{5pt}
 \setlength{\belowdisplayskip}{5pt}
 \setlength{\abovedisplayshortskip}{5pt}
 \setlength{\belowdisplayshortskip}{5pt}
\begin{equation}
    \widehat{hdg}_t = \operatorname{atan2}(\dot{y}_t,\dot{x}_t),
    \label{eq:heading}
\end{equation}}
or, when velocities are not part of the state, from successive predicted
positions:
{\setlength{\abovedisplayskip}{5pt}
 \setlength{\belowdisplayskip}{5pt}
 \setlength{\abovedisplayshortskip}{5pt}
 \setlength{\belowdisplayshortskip}{5pt}
\begin{equation}
\widehat{hdg}_t=\operatorname{atan2}(y_t-y_{t-1},\,x_t-x_{t-1}).
\end{equation}}
Given the fixed radar position $(x_r,y_r)$, the line-of-sight (LOS) azimuth from
the radar to the target is
{\setlength{\abovedisplayskip}{5pt}
 \setlength{\belowdisplayskip}{5pt}
 \setlength{\abovedisplayshortskip}{5pt}
 \setlength{\belowdisplayshortskip}{5pt}
\begin{equation}
\widehat{\theta}_t=\operatorname{atan2}(y_t-y_r,\,x_t-x_r).
\label{eq:los_azimuth}
\end{equation}}
Following Sec.~\ref{subsec:HRRPdata}, we estimate the aspect angle as the
difference between heading and LOS azimuth and wrap it to $[0,2\pi)$:
{\setlength{\abovedisplayskip}{-3pt}
 \setlength{\belowdisplayskip}{5pt}
 \setlength{\abovedisplayshortskip}{-3pt}
 \setlength{\belowdisplayshortskip}{5pt}
\begin{equation}
    \widehat{\asp}_t = \rm{wrap}_{[0,2 \pi)}(\widehat{hdg}_t - \widehat{\theta}_t) ,    \label{eq:aspect}
\end{equation}}
where $\mathrm{wrap}_{[0,2\pi)}(\theta)
=
\theta - 2\pi \left\lfloor \frac{\theta}{2\pi} \right\rfloor$ returns values in $[0,2\pi)$. The Kalman predictor
thus provides both a smoothed kinematic trajectory and the aspect-angle
estimates used for conditioning. In our experiments, we split trajectories into
segments when the gap between two consecutive measurements exceeds 20 minutes.
Kalman parameters are tuned to best match the statistics of our measured
trajectories. Kalman estimates are computed online, using only past and 
current measurements (causal).

\subsection{Conditioning Methods}
\label{subsec:conditioning_methods}
To incorporate aspect-angle information, we investigate three conditioning
strategies commonly used in the literature: concatenation, Feature-wise Linear
Modulation (FiLM), and Conditional Batch Normalization (CBN). Let
$x \in \mathbb{R}^{N \times C \times L}$ denote an intermediate feature tensor
(batch size $N$, $C$ channels, length $L$), and let $c \in \mathbb{R}^{N \times D}$
be the associated conditioning vector ($D$-dimensional angle encoding).

\subsubsection{Concatenation}
Concatenation expands $c$ to match the spatial support of $x$ and appends it
along the channel dimension. We map $c$ to a scalar token per sample using a
linear projection $g_\phi:\mathbb{R}^{D}\to\mathbb{R}$, reshape it as $(N,1,1)$,
and broadcast it along the length axis to obtain $(N,1,L)$. Concatenating with
$x$ yields an augmented feature map of shape $(N,C+1,L)$.

\subsubsection{Feature-wise Linear Modulation (FiLM)}
\cite{film}
FiLM conditions $x$ through a per-channel affine modulation. For each sample
$n$, a learnable predictor $f_\theta$ (implemented as a linear layer) outputs a
scale and shift from $c_n$:
{\setlength{\abovedisplayskip}{5pt}
 \setlength{\belowdisplayskip}{5pt}
 \setlength{\abovedisplayshortskip}{5pt}
 \setlength{\belowdisplayshortskip}{5pt}
\begin{equation}
(\gamma(c_n),\,\beta(c_n)) = f_\theta(c_n),
\label{eq:film_affine_predictor}
\end{equation}}
where $\gamma(c_n),\beta(c_n)\in\mathbb{R}^{C}$. The FiLM output is
{\setlength{\abovedisplayskip}{5pt}
 \setlength{\belowdisplayskip}{5pt}
 \setlength{\abovedisplayshortskip}{5pt}
 \setlength{\belowdisplayshortskip}{5pt}
\begin{equation}
y_{n,c,l}=\gamma_c(c_n)\,x_{n,c,l}+\beta_c(c_n),
\label{eq:film}
\end{equation}}
with broadcasting over the length index $l$.

\subsubsection{Conditional Batch Normalization (CBN)}
\cite{cbn}
CBN replaces the standard BatchNorm affine parameters with sample-dependent
parameters predicted from $c$. First, BatchNorm computes channel-wise
normalized activations:
\begin{equation}
\widehat{x}_{n,c,l}=\frac{x_{n,c,l}-\mu_c}{\sqrt{\sigma_c^2+\epsilon}},
\label{eq:bn_norm}
\end{equation}
where $\mu_c$ and $\sigma_c^2$ are the batch mean and variance for channel $c$
(computed over indices $(n,l)$), and $\epsilon>0$ is a small constant. In CBN,
the affine parameters are predicted from the conditioning input:
{\setlength{\abovedisplayskip}{5pt}
 \setlength{\belowdisplayskip}{5pt}
 \setlength{\abovedisplayshortskip}{5pt}
 \setlength{\belowdisplayshortskip}{5pt}
\begin{equation}
(\gamma(c_n),\,\beta(c_n)) = f_\theta(c_n),
\label{eq:cbn_affine_predictor}
\end{equation}}
with $\gamma(c_n),\beta(c_n)\in\mathbb{R}^{C}$. The output is
{\setlength{\abovedisplayskip}{5pt}
 \setlength{\belowdisplayskip}{5pt}
 \setlength{\abovedisplayshortskip}{5pt}
 \setlength{\belowdisplayshortskip}{5pt}
\begin{equation}
y_{n,c,l}=\gamma_c(c_n)\,\widehat{x}_{n,c,l}+\beta_c(c_n).
\label{eq:cbn}
\end{equation}}

FiLM applies a sample-dependent affine transform directly to activations,
whereas CBN first normalizes activations using batch statistics and then
modulates them via sample-dependent affine parameters.

\section{Results}

\subsection{Datasets}
We evaluate aspect-angle conditioning on three datasets: an HRRP version of MSTAR and two measured ship
HRRP datasets.

\paragraph{MSTAR-HRRP}
We provide a reproducible processing pipeline for the publicly available MSTAR Phoenix SAR chips. It
reads the magnitude/phase files, reconstructs the complex SAR image, applies an inverse transform along
cross-range, aggregates amplitude returns into a 128-cell HRRP, and stores aspect/depression metadata.
The resulting dataset contains 10 military vehicle classes with aspect angles spanning
$[0^\circ,360^\circ)$ at a fixed depression angle.

\paragraph{Ship datasets}
Both ship datasets are built from a measured maritime HRRP database, which provides time stamps, 
AIS-based kinematics, and an AIS-heading-based reference aspect angle (AIS + radar geometry) for each profile. We define 
two subsets, denoted Ship (A) and Ship (B), to study angle conditioning under different levels 
of class imbalance and inter-class ambiguity.

\begin{wrapfigure}[11]{l}{0.45\linewidth}
    \centering
    \vspace{-10pt}
    \includegraphics[width=1.\linewidth]{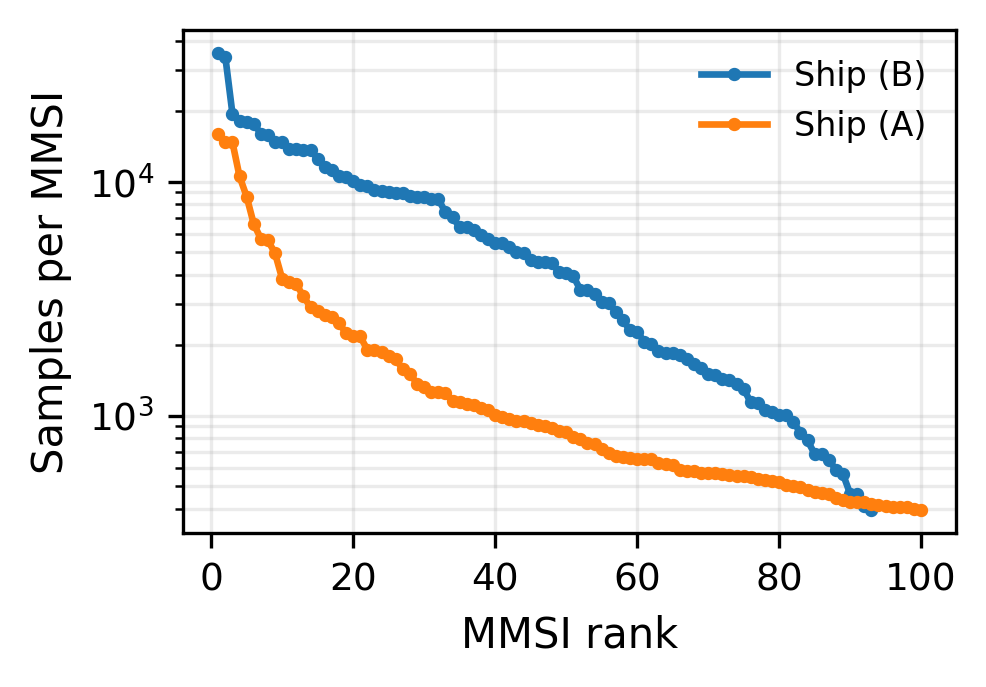}
    \caption{Sorted MMSI class frequencies for Ship (A) and Ship (B) datasets.}
    \label{fig:mmsi_compact}
\end{wrapfigure}

Classes are defined by the ship MMSI (Maritime Mobile Service Identity), i.e., the unique 9-digit identifier
assigned to each vessel (100 classes for Ship (A) and 93 for Ship (B)). Fig.~\ref{fig:mmsi_compact} indicates 
that Ship (A) is more class-imbalanced than Ship (B). Moreover, Ship (B) provides more uniform aspect-angle coverage, 
making it a more reliable setting to evaluate angle-aware learning with fewer angle--class confounds.

Ship (A) includes 100 ships and 185k HRRP profiles. It is intentionally ambiguous: ships exhibit only
24 distinct lengths (mean $\approx$ 100\,m) and five distinct widths, leading to many classes with similar
macroscopic geometry. Ship (B) includes 93 ships and about 600k profiles, with more diverse dimensions
(12--400\,m length) and better coverage of the full $360^\circ$ aspect-angle range.

\subsection{Aspect-angle estimation quality}
In practice, aspect angles are not directly measured and must be estimated from kinematic information.
Since we later report results with predicted angles (\emph{Pred aspect}), we first quantify the accuracy
of the position-only estimator described in Sec.~\ref{subsec:aspect_angle_estimation} (Eqs.~(\ref{eq:heading})\, \hspace{-7pt}
--\,(\ref{eq:aspect})).

\begin{wrapfigure}[17]{r}{0.5\linewidth}
    \centering
    \vspace{-17pt}
    \includegraphics[width=\linewidth]{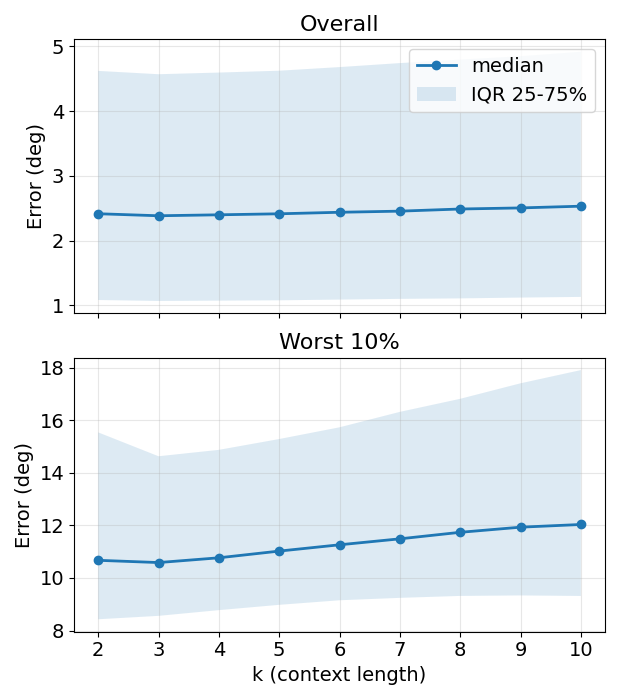}
    \caption{Position-only aspect-angle estimation error over 100k segments (all segments and worst 10\%).}
    \label{fig:asp_est}
\end{wrapfigure}

To this end, we sample 100k contiguous trajectory segments and evaluate the wrapped angular error with
respect to the reference aspect angle over increasing context lengths $k\in\{2,\dots,10\}$. We average
errors over $k$ to obtain a segment-level score and also analyze the worst 10\% segments to highlight
typical failure cases.

Fig.~\ref{fig:asp_est} shows a median error below $3^\circ$ over all segments, while the worst 10\%
remain around $11^\circ$--$12^\circ$. The interquartile range is stable across context lengths, indicating that
the simple causal estimator is accurate enough for the following experiments.

\newcommand{\accmf}[2]{\shortstack{#1\\#2}}
\renewcommand{\arraystretch}{1.15}
\newcommand{\dsfour}[1]{\multirow{4}{*}{\raisebox{-3.7ex}[0pt][0pt]{#1}}}
\newcommand{\dsthree}[1]{\multirow{3}{*}{\raisebox{-3.2ex}[0pt][0pt]{#1}}}

\begin{table*}[!t]
    \centering
    \caption{One-view classification results with different conditioning methods and architectures (Accuracy \textbar \hspace{1pt} Macro F1 [\%]).}
    \label{tab:real_aspect_one_view}
    \begingroup
    \footnotesize
    \setlength{\tabcolsep}{3pt}
    \renewcommand{\arraystretch}{0.95}
    \renewcommand{\accmf}[2]{#1\,\textbar\,#2}
    \begin{adjustbox}{max width=\textwidth}
    \begin{tabular}{@{}c*{9}{c}@{}}
        \toprule
        \multirow{2}{*}{\textbf{Conds}} & \multicolumn{3}{c}{\textbf{MSTAR}} & \multicolumn{3}{c}{\textbf{Ship (A)}} & \multicolumn{3}{c}{\textbf{Ship (B)}} \\
        \cmidrule(lr){2-4}\cmidrule(lr){5-7}\cmidrule(lr){8-10}
        & \textbf{ResNet} & \textbf{MLP} & \textbf{Conv} & \textbf{ResNet} & \textbf{MLP} & \textbf{Conv} & \textbf{ResNet} & \textbf{MLP} & \textbf{Conv} \\
        \midrule
        Uncond & \accmf{89.46}{89.27} & \accmf{90.34}{89.84} & \accmf{74.27}{73.47} & \accmf{64.29}{59.71} & \accmf{27.47}{21.60} & \accmf{60.36}{53.60} & \accmf{73.62}{78.29} & \accmf{37.48}{34.05} & \accmf{69.88}{72.25} \\
        \cmidrule(lr){1-10}
        Concat & \accmf{92.37}{91.86} & \accmf{96.08}{95.80} & \accmf{\textbf{83.78}}{\textbf{83.05}} & \accmf{\textbf{73.56}}{68.08} & \accmf{\textbf{33.24}}{\textbf{26.23}} & \accmf{69.49}{63.30} & \accmf{78.45}{83.32} & \accmf{\textbf{43.28}}{39.86} & \accmf{68.20}{70.71} \\
        \cmidrule(lr){1-10}
        FiLM & \accmf{93.60}{93.19} & \accmf{95.42}{95.32} & \accmf{80.01}{79.11} & \accmf{73.28}{\textbf{68.14}} & \accmf{32.56}{26.18} & \accmf{\textbf{69.62}}{\textbf{63.48}} & \accmf{78.61}{83.73} & \accmf{42.81}{\textbf{40.04}} & \accmf{\textbf{76.39}}{\textbf{80.42}} \\
        \cmidrule(lr){1-10}
        CBN & \accmf{\textbf{95.21}}{\textbf{95.05}} & \accmf{\textbf{96.23}}{\textbf{96.08}} & \accmf{80.30}{79.73} & \accmf{69.35}{63.38} & \accmf{30.34}{23.81} & \accmf{64.69}{57.51} & \accmf{\textbf{79.02}}{\textbf{84.34}} & \accmf{42.81}{\textbf{40.04}} & \accmf{76.06}{80.19} \\
        \bottomrule
    \end{tabular}
    \end{adjustbox}
    \endgroup
\end{table*}

\subsection{Experimental setup}

To reduce overfitting to exact aspect-angle values, we add Gaussian noise to the conditioning angle
 ($\sigma\!=\!2^\circ$); no jitter is used at validation/test time.

\subsubsection{One-view classification}
For one-view classification, we split each dataset into training, validation, and test sets with a
70\%/15\%/15\% ratio using a label-stratified split. We train our three architectures with each 
conditioning method and compare against an unconditioned baseline.

\subsubsection{Multi-view classification}
For multi-view classification, the split requires additional care because adjacent aspect angles tend
to cluster within the same time windows. A naive chronological split can therefore induce a shift in
aspect-angle distributions across train/validation/test sets, which hinders generalization.

We sort each ship’s profiles by acquisition time, split them into train/val/test (70/15/15), and build
sequences by grouping profiles from the same ship within the same split. A strictly contiguous time
split would strongly skew aspect-angle coverage. Our split instead prevents cross-set overlap while
keeping angle distributions more comparable, at the cost of less realistic (non-contiguous) sequences
since angles along real trajectories are typically more correlated. We only evaluate multi-view models
on ships: although MSTAR provides multi-aspect measurements, it does not form operational trajectories,
so sequences would be artificial and not meaningful for view aggregation.

\subsection{Results}

\subsubsection{One-view classification}
Table~\ref{tab:real_aspect_one_view} reports one-view results. Across datasets and architectures, injecting
aspect-angle information improves both accuracy and macro-F1 compared to unconditioned baselines, confirming
the benefit of angle awareness for HRRP classification. ResNet consistently achieves the strongest overall
performance, motivating its use as backbone in multi-view experiments. FiLM and CBN yield comparable results
in most settings; we therefore select CBN in the sequential study.

Table~\ref{tab:aspect_estimation} evaluates conditioning with the estimated aspect angle on the ResNet backbone. 
Using position-only estimated angles yields lower one-view performance than reference angles, but remains
consistently above the unconditioned baseline on both ship datasets. Performance is also stable across
context lengths $k$, showing that the estimated angles preserve a useful geometric cue despite residual
estimation errors.

\begin{table}[!htbp]
    \centering
    \caption{One-view classification with estimated aspect angles on Ship (A)/(B) (Acc./Macro-F1).}
    \label{tab:aspect_estimation}
    \resizebox{\linewidth}{!}{%
    \begin{tabular}{@{}ccccc@{}}
        \toprule
        Dataset & Conds & \shortstack{k=2\\(Acc / Macro F1)} & \shortstack{k=5\\(Acc / Macro F1)} & \shortstack{k=10\\(Acc / Macro F1)} \\
        \midrule
        \multirow{2}{*}{Ship (A)} & Uncond & \accmf{64.29\%$\pm$0.45\%}{59.71\%$\pm$0.70\%} & \accmf{64.21\%$\pm$0.61\%}{59.33\%$\pm$0.79\%} & \accmf{64.41\%$\pm$0.42\%}{59.54\%$\pm$0.68\%} \\
        \cmidrule(lr){2-5}
        & CBN & \accmf{\textbf{66.62\%$\pm$0.57\%}}{\textbf{62.85\%$\pm$0.75\%}} & \accmf{\textbf{67.74\%$\pm$1.24\%}}{\textbf{64.14\%$\pm$1.33\%}} & \accmf{\textbf{67.24\%$\pm$1.50\%}}{\textbf{64.10\%$\pm$1.23\%}} \\
        \midrule
        \multirow{2}{*}{Ship (B)} & Uncond & \accmf{73.72\%$\pm$0.59\%}{78.38\%$\pm$0.71\%} & \accmf{74.01\%$\pm$0.60\%}{78.74\%$\pm$0.88\%} & \accmf{74.02\%$\pm$0.82\%}{78.70\%$\pm$1.19\%} \\
        \cmidrule(lr){2-5}
        & CBN & \accmf{\textbf{75.85\%$\pm$0.48\%}}{\textbf{80.51\%$\pm$0.72\%}} & \accmf{\textbf{76.35\%$\pm$0.57\%}}{\textbf{80.71\%$\pm$0.50\%}} & \accmf{\textbf{77.02\%$\pm$0.63\%}}{\textbf{81.50\%$\pm$0.49\%}} \\
        \bottomrule
    \end{tabular}
    }
\end{table}

\subsubsection{Multi-view classification}

Table~\ref{tab:two_entry} reports multi-view results on Ship (A) and Ship (B). Without angle input,
performance varies widely across sequence models and can remain low on Ship (A), whereas providing
aspect angles consistently yields strong accuracy and macro-F1. This gap is smaller on Ship (B), which
exhibits more diverse ship dimensions and a more uniform aspect-angle coverage.

\begin{table}[!htbp]
    \centering
    \caption{Multi-view results on ship datasets (A) and (B) (Accuracy / Macro F1).}
    \label{tab:two_entry}
    \resizebox{\linewidth}{!}{%
    \begin{tabular}{@{}ccccc@{}}
        \toprule
        \textbf{Dataset} & \textbf{Conds} & \shortstack{\textbf{LSTM}\\\textbf{(Acc / Macro F1)}} & \shortstack{\textbf{GRU}\\\textbf{(Acc / Macro F1)}} & \shortstack{\textbf{Transformer}\\\textbf{(Acc / Macro F1)}} \\
        \midrule
        \dsthree{Ship (A)}
            & None & \accmf{43.14\%$\pm$17.62\%}{39.63\%$\pm$19.00\%} & \accmf{42.26\%$\pm$6.36\%}{41.23\%$\pm$4.65\%} & \accmf{65.33\%$\pm$3.50\%}{66.08\%$\pm$4.24\%} \\
        \cmidrule(lr){2-5}
            & Real aspect & \accmf{\textbf{93.88\%$\pm$0.16\%}}{92.40\%$\pm$0.64\%} & \accmf{\textbf{94.34\%$\pm$1.49\%}}{92.99\%$\pm$1.93\%} & \accmf{95.54\%$\pm$4.48\%}{94.72\%$\pm$5.54\%} \\
        \cmidrule(lr){2-5}
            & Pred aspect & \accmf{92.83\%$\pm$0.99\%}{\textbf{92.46\%$\pm$1.29\%}} & \accmf{93.65\%$\pm$1.51\%}{\textbf{93.80\%$\pm$1.36\%}} & \accmf{\textbf{96.95\%$\pm$0.64\%}}{\textbf{96.93\%$\pm$0.51\%}} \\
        \midrule
        \dsthree{Ship (B)}
            & None & \accmf{80.43\%$\pm$1.44\%}{88.08\%$\pm$1.70\%} & \accmf{80.27\%$\pm$1.43\%}{88.14\%$\pm$1.57\%} & \accmf{79.95\%$\pm$2.01\%}{87.50\%$\pm$2.71\%} \\
        \cmidrule(lr){2-5}
            & Real aspect & \accmf{87.28\%$\pm$3.66\%}{93.00\%$\pm$2.45\%} & \accmf{\textbf{88.31\%$\pm$1.13\%}}{\textbf{93.96\%$\pm$1.37\%}} & \accmf{\textbf{90.59\%$\pm$0.89\%}}{\textbf{96.07\%$\pm$0.69\%}} \\
        \cmidrule(lr){2-5}
            & Pred aspect & \accmf{\textbf{88.08\%$\pm$0.77\%}}{\textbf{93.75\%$\pm$0.44\%}} & \accmf{87.35\%$\pm$1.56\%}{93.01\%$\pm$1.19\%} & \accmf{90.18\%$\pm$0.82\%}{95.63\%$\pm$0.68\%} \\
        \bottomrule
    \end{tabular}
    }
\end{table}

Using position-only estimated angles (\emph{Pred aspect}) achieves performance close to using reference angles
(\emph{Real aspect}) in the multi-view setting, with differences within run-to-run variability. Finally, 
using CBN with estimated angles remains stable in our experiments, suggesting robustness to
moderate conditioning noise.

\section{Discussion}
Aspect-angle conditioning improves HRRP classification across datasets, and multi-view aggregation
provides additional gains by combining complementary viewpoints. Predicted angles (\emph{Pred aspect})
perform close to reference angles in multi-view experiments. In one-view classification they remain below
reference-angle conditioning, but still clearly improve over unconditioned models, confirming that practical
angle estimates retain discriminative information.

A key caveat is that angle coverage may differ across vessels: in Ship~(A), angle-conditioned models can
reach unexpectedly high performance despite strong inter-class ambiguity, suggesting that dataset-specific
angle patterns may act as a shortcut cue. In contrast, Ship~(B), with more uniform angular coverage,
offers a more reliable assessment of the intrinsic benefit of angle awareness.

\section{Conclusion}
We studied the impact of aspect-angle awareness for HRRP classification in both one-view and multi-view
settings, using an HRRP version of MSTAR and two measured maritime datasets. Across architectures and
datasets, injecting aspect-angle information consistently improves accuracy and macro-F1, highlighting
the central role of acquisition geometry in shaping 1D range signatures. Multi-view aggregation further
boosts performance by combining complementary viewpoints along a trajectory.

We also evaluated angle-aware models under imperfect angle inputs using a Kalman-based online estimator.
Overall, conditioning on predicted angles yields performance close to using reference angles in the
multi-view setting, with a slightly larger gap in one-view classification, supporting the feasibility of
angle-conditioned recognition with realistic angle estimates. Finally, our ship experiments underline
that angle conditioning can interact with dataset-specific angle coverage; careful control of
angle--class correlations is therefore important to obtain unbiased evaluations of multi-view models.

\bibliographystyle{IEEEbib}
\bibliography{biblio}

\begin{thebibliography}{10}

\bibitem{Wan2019ConvolutionalNN}
J.~Wan, B.~Chen, B.~Xu, H.~Liu, and L.~Jin,
\newblock ``Convolutional neural networks for radar hrrp target recognition and
  rejection,''
\newblock {\em EURASIP J. Adv. Signal Process.}, vol. 2019, pp. 5, 2019.

\bibitem{bauw2020unsupervised}
M.~Bauw, S.~Velasco-Forero, J.~Angulo, C.~Adnet, and O.~Airiau,
\newblock ``From unsupervised to semi-supervised anomaly detection methods for
  hrrp targets,''
\newblock {\em 2020 IEEE Radar Conference (RadarConf20)}, pp. 1--6, 2020.

\bibitem{Sun}
L.~Sun, J.~Liu, Y.~Liu, and B.~Li,
\newblock ``Hrrp target recognition based on soft-boundary deep svdd with
  lstm,''
\newblock in {\em Proc. Int. Conf. Control, Autom. Inf. Sci. (ICCAIS)}, 2021,
  pp. 1047--1052.

\bibitem{hrrp_clf1}
Y.~Diao, S.~Liu, X.~Gao, A.~Liu, and Z.~Zhang,
\newblock ``Cnn based on multiscale window self-attention mechanism for radar
  hrrp target recognition,''
\newblock in {\em 2022 7th International Conference on Signal and Image
  Processing (ICSIP)}, 2022, pp. 281--285.

\bibitem{hrrp_seq1}
C.-L. Lin, T.-P. Chen, K.-C. Fan, H.-Y. Cheng, and C.-H. Chuang,
\newblock ``Radar high-resolution range profile ship recognition using
  two-channel convolutional neural networks concatenated with bidirectional
  long short-term memory,''
\newblock {\em Remote Sensing}, vol. 13, no. 7, 2021.

\bibitem{hrrp_seq2}
B.~Xu, B.~Chen, J.~Wan, H.~Liu, and L.~Jin,
\newblock ``Target-aware recurrent attentional network for radar {HRRP} target
  recognition,''
\newblock {\em Signal Processing}, vol. 155, pp. 268--280, 2019.

\bibitem{hrrp_seq3}
X.~Wang, P.~Wang, Y.~Song, Q.~Xiang, and J.~Li,
\newblock ``Recognition of high-resolution range profile sequence based on tcn
  with sequence length-adaptive algorithm and elastic net regularization,''
\newblock {\em Expert Syst. Appl.}, vol. 248, pp. 123417, 2024.

\bibitem{multi_aspect_gen}
Y.~Song, Q.~Zhou, W.~Yang, Y.~Wang, C.~Hu, and X.~Hu,
\newblock ``Multi-view hrrp generation with aspect-directed attention gan,''
\newblock {\em IEEE Journal of Selected Topics in Applied Earth Observations
  and Remote Sensing}, vol. 15, pp. 7643--7656, 2022.

\bibitem{mfn}
E.~Brient, S.~Velasco-Forero, and R.~Kassab,
\newblock ``Mfn decomposition and related metrics for high-resolution range
  profiles generative models,''
\newblock in {\em Proc. IEEE Radar Conf. (RadarConf)}. IEEE, 2025, pp. 1--6.

\bibitem{domain_adapt_hrrp}
Y.~Wang, Y.~Ma, L.~Zhang, J.~Wang, Y.~Zhang, and H.~Lv,
\newblock ``Disentanglement model for hrrp target recognition when missing
  aspects,''
\newblock in {\em Proc. IEEE Int. Geosci. Remote Sens. Symp. (IGARSS)}, 2023,
  pp. 5770--5773.

\bibitem{domain_adapt_hrrp2}
Y.~Wen, L.~Shi, X.~Yu, Y.~Huang, and X.~Ding,
\newblock ``Hrrp target recognition with deep transfer learning,''
\newblock {\em IEEE Access}, vol. 8, pp. 57859--57867, 2020.

\bibitem{constrastive_learning_ma}
Y.~Zhong, W.~Lin, Y.~Xu, L.~Huang, Y.~Huang, and X.~Ding,
\newblock ``Contrastive learning for radar hrrp recognition with missing
  aspects,''
\newblock {\em IEEE Geoscience and Remote Sensing Letters}, vol. 20, pp. 1--5,
  2023.

\bibitem{richards2005fundamentals}
M.~Richards,
\newblock {\em Fundamentals Of Radar Signal Processing},
\newblock McGraw-Hill Education (India) Pvt Limited, 2005.

\bibitem{resnet}
K.~He, X.~Zhang, S.~Ren, and J.~Sun,
\newblock ``Deep residual learning for image recognition,''
\newblock in {\em Proc. IEEE Conf. Comput. Vis. Pattern Recognit. (CVPR)},
  2016, pp. 770--778.

\bibitem{welch1995introduction}
G.~Welch, G.~Bishop, and C.~Hill,
\newblock ``{An introduction to the Kalman filter},''
\newblock pp. 1--16, 1995.

\bibitem{film}
E.~Perez, F.~Strub, H.~de~Vries, V.~Dumoulin, and A.~C. Courville,
\newblock ``Film: Visual reasoning with a general conditioning layer,''
\newblock in {\em AAAI}, 2018.

\bibitem{cbn}
H.~de~Vries, F.~Strub, J.~Mary, H.~Larochelle, O.~Pietquin, and A.~C.
  Courville,
\newblock ``Modulating early visual processing by language,''
\newblock in {\em Adv. Neural Inf. Process. Syst.}, I.~Guyon, U.~V. Luxburg,
  S.~Bengio, H.~Wallach, R.~Fergus, S.~Vishwanathan, and R.~Garnett, Eds. 2017,
  vol.~30, Curran Associates, Inc.

\end{thebibliography}

\end{document}